\documentclass[twocolumn]{jpsj3} 

\usepackage{color}

\title{
Charge, Lattice, and Spin Dynamics in Photoinduced Phase Transitions from Charge-Order-Insulator to Metal in Quasi-Two-Dimensional Organic Conductors 
}

\author{
Satoshi \textsc{Miyashita}$^{1,2}$, 
Yasuhiro \textsc{Tanaka}$^{2}$, 
Shinichiro \textsc{Iwai}$^{3,4}$, and 
Kenji \textsc{Yonemitsu}$^{2,5}$\thanks{E-mail: kxy@ims.ac.jp}
}

\inst{
$^{1}$Institute for Materials Research, Tohoku University, Sendai, Miyagi 980-8577, Japan \\
$^{2}$Institute for Molecular Science, Okazaki, Aichi 444-8585, Japan \\
$^{3}$Department of Physics, Tohoku University, Sendai, Miyagi 980-8578, Japan \\
$^{4}$JST, CREST, Sendai, Miyagi 980-8578, Japan \\
$^{5}$Department of Functional Molecular Science, Graduate University for Advanced Studies, Okazaki, Aichi 444-8585, Japan
}

\abst{
To elucidate different photoinduced melting dynamics of charge orders observed in quasi-two-dimensional organic conductors $ \theta $-(BEDT-TTF)$_2$RbZn(SCN)$_4$ and $ \alpha $-(BEDT-TTF)$_2$I$_3$ [BEDT-TTF=bis(ethylenedithio)tetrathiafulvalene], we theoretically study photoinduced time evolution of charge and spin correlation functions on the basis of exact many-electron wave functions coupled with classical phonons in extended Peierls-Hubbard models on anisotropic triangular lattices. In both salts, the so-called horizontal-stripe charge order is stabilized by nearest-neighbor repulsive interactions and by 
electron-lattice interactions. In $ \theta $-(BEDT-TTF)$_2$RbZn(SCN)$_4$ (abbreviated as $ \theta $-RbZn), the stabilization energy due to lattice distortion is larger, so that larger quantity of energy needs to be absorbed for the melting of the charge and lattice orders. The photoinduced charge dynamics shows a complex behavior owing to a substantial number of nearly degenerate eigenstates involved. This is related to the high structural symmetry when the lattice is undistorted. In $ \alpha $-(BEDT-TTF)$_2$I$_3$ (abbreviated as $ \alpha $-I$_3$), the lattice stabilization energy is smaller, and smaller quantity of energy is sufficient to melt the charge and lattice orders leading to a metallic phase. The photoinduced charge dynamics shows a sinusoidal oscillation. In $ \alpha $-I$_3$, the low structural symmetry ensures nearly spin-singlet bonds between hole-rich sites, where the spin correlation survives even after photoexcitation. 
}

\kword{
photoinduced phase transition, charge order, metal-insulator transition, frustration
}

\begin{document}
\maketitle

\section{Introduction}

Electronic phases are controlled by changing pressure, constituents, or temperature in and near equilibrium conditions. The effects of changing pressure or constituents can be introduced to theoretical calculations by changing model parameters. Electronic phases are dynamically controlled in nonequilibrium conditions as well, where the control process cannot be described by simply changing model parameters. 
Among them, photoinduced phase transitions have attracted much attention. \cite{yonemitsu_jpsj06,yonemitsu_pr08} Along with the development of time-resolved spectroscopy, ultrafast transition dynamics enabled by strong electron correlation is one of the key issues at present. 

A considerable number of ultrafast photoinduced transitions toward metallic phases have been observed in molecular materials. Transitions from Mott-insulator phases are realized in a halogen-bridged nickel-chain compound \cite{iwai_prl03,ono_prb04} and in a quasi-one-dimensional half-filled-band organic salt (BEDT-TTF)(F$_2$TCNQ) \cite{okamoto_prl07} (F$_2$TCNQ=2,5-difluorotetracyanoquinodimethane). They are theoretically studied in refs.~\citen{maeshima_jpsj05,maeshima_jpcs05,jdlee_prb07,takahashi_prb08}. Ultrafast photoinduced transitions in charge density wave, Mott insulator, and metallic phases are investigated in an iodine-bridged platinum-chain compound. \cite{kimura_prb09} Transitions from charge-ordered-insulator phases are realized in a quasi-one-dimensional quarter-filled-band organic salt (EDO-TTF)$_2$PF$_6$ \cite{chollet_s05,onda_prl08} (EDO-TTF=ethylenedioxy-tetrathiafulvalene) and in quasi-two-dimensional quarter-filled-band organic salts $ \theta $-(BEDT-TTF)$_2$RbZn(SCN)$_4$ \cite{iwai_prl07} ($ \theta $-RbZn) and $ \alpha $-(BEDT-TTF)$_2$I$_3$ \cite{iwai_prl07,iwai_prb08} ($ \alpha $-I$_3$). The photoinduced dynamics in the former salt is also theoretically studied. \cite{yonemitsu_prb07,onda_prl08} The ground states of the latter salts are theoretically shown to be stabilized by nearest-neighbor repulsive interactions and by electron-lattice interactions. \cite{tanaka_jpsj07,miyashita_prb07,tanaka_jpsj08,miyashita_jpsj08,tanaka_jpsj09}

The charge ordering in quasi-two-dimensional organic systems has been intensively studied both experimentally \cite{mori_prb98,dressel_cr04,takahashi_jpsj06} and theoretically. \cite{seo_cr04,seo_jpsj06r} The low-temperature phases are shown to have horizontal-stripe charge orders by NMR experiments (for $ \theta $-RbZn \cite{miyagawa_prb00,chiba_jpcs01} and for 
$ \alpha $-I$_3$ \cite{takano_jpcs01,takano_sm01}) and others. Their crystal structures are well analyzed by X-ray diffraction studies (for $ \theta $-RbZn \cite{watanabe_jpsj04} and for 
$ \alpha $-I$_3$ \cite{kakiuchi_jpsj07}). 
\begin{figure}
\includegraphics[height=17cm]{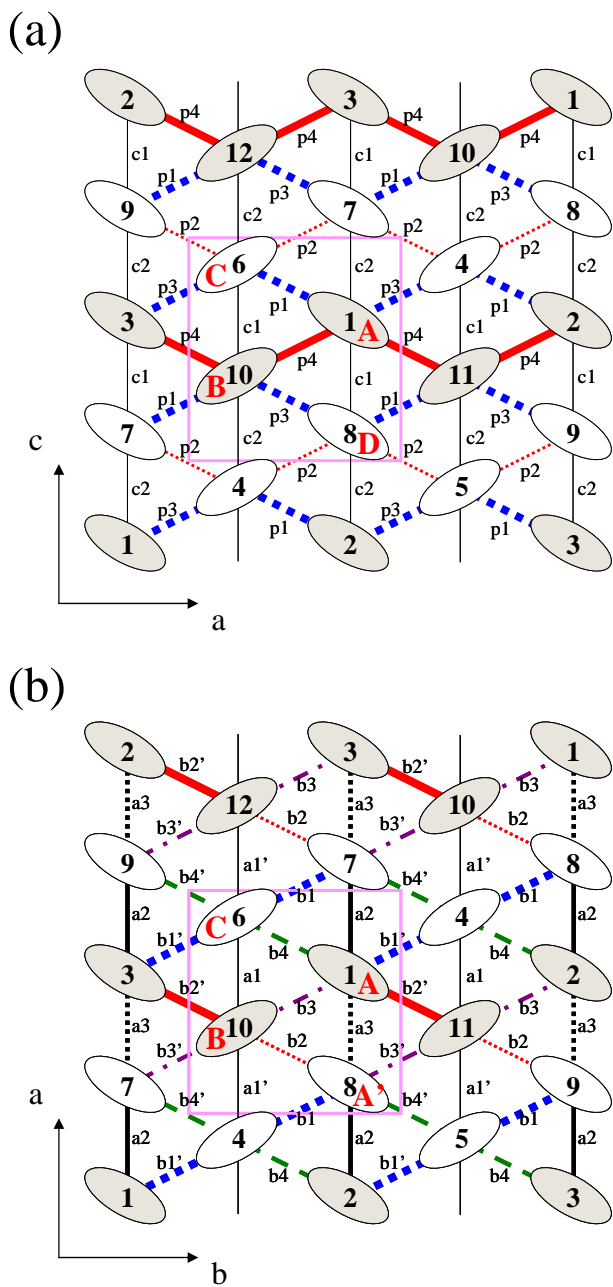}
\caption{(Color online) Anisotropic triangular lattices for (a) $ \theta $-RbZn and (b) $ \alpha $-I$_3$. The rectangles denote the unit cells. The gray and white ellipses represent the hole-rich and hole-poor sites, respectively. The red or thickest lines possess the largest absolute values of transfer integrals after electron-lattice couplings are taken into account. \label{fig:structure}}
\end{figure}
Figure~\ref{fig:structure} schematically shows the low-temperature structures of the conduction layers with charge orders. The hole-rich sites indicated by the gray ellipses are aligned horizontally in this figure. For the optical properties of these salts, see ref.~\citen{dressel_cr04} for example. For $ \theta $-(BEDT-TTF)$_2$CsZn(SCN)$_4$ (where Rb in $ \theta $-RbZn is replaced by Cs), which does not have a long-range order even in the nonmetallic phase at low temperatures, charge frustration effects are suggested to be important by permittivity measurements \cite{inagaki_jpsj04} and by 
the observation of thyristor behavior. \cite{sawano_n05} In  $ \theta $-RbZn, lattice effects are discussed for the conductance and dielectric permittivity at various cooling and heating rates. \cite{nad_prb07} In $ \alpha $-I$_3$, ferroelectric polarization due to the charge ordering is recently reported, \cite{yamamoto_jpsj08} 
where the ferroelectric order is of almost purely electronic origin. 

After the pioneering theoretical work for the electronic states of quasi-two-dimensional organic conductors, \cite{kino_jpsj96} the charge-ordered states have been studied mainly in extended Hubbard models because the nearest-neighbor repulsive interactions mainly stabilize them. \cite{mckenzie_s97,seo_jpsj00,mckenzie_prb01,clay_jpsj02,calandra_prb02,merino_prb03,kobayashi_jpsj04,merino_prl06} Charge frustration effects are recognized in various aspects. \cite{merino_prb05,kaneko_jpsj06,watanabe_jpsj06,seo_jpsj06} Sometimes they are discussed in spinless fermion models, \cite{hotta_prb06,hotta_jpsj06} but the spin degrees of freedom are pointed out as important for determining the relative stability of different charge ordering patterns. \cite{miyashita_jpsj08} Recent theories for charge orderings include refs.~\citen{kuroki_jpsj06,udagawa_prl07,nishimoto_prb08,mazumdar_prb08,merino_prb09}. In such frustrated systems, the lattice degrees of freedom are also important, \cite{tanaka_jpsj07,miyashita_prb07,tanaka_jpsj08,miyashita_jpsj08,tanaka_jpsj09} so that extended Peierls-Hubbard models on anisotropic triangular lattices are studied in the present work. 

As mentioned above, photoinduced melting of the charge orders has been observed in these salts by femtosecond reflection spectroscopy. \cite{iwai_prl07,iwai_prb08} It has revealed a fundamental difference between their dynamics, which would be due to different stabilities of these charge orders. The $ \theta $-RbZn salt shows local melting and ultrafast recovery to the charge order in 0.2 ps irrespective of temperature and of 
excitation intensity. The $ \alpha $-I$_3$ salt shows critical slowing down in the transient reflectivity spectra, where decay times depend largely on temperature and on 
excitation intensity and vary from the picosecond timescale to the ten-nanosecond timescale. The relative importance of nearest-neighbor repulsive interactions and electron-lattice interactions is shown to be different between these two salts in our previous studies using the unrestricted Hartree-Fock approximation, \cite{tanaka_jpsj08} the exact diagonalization, and the strong-coupling perturbation theory. \cite{miyashita_jpsj08} 

In $ \theta $-RbZn, the lattice effect is strong and required to stabilize the horizontal-stripe charge order among different patterns of charge orders that are nearly degenerate if the lattice is undistorted and has the high structural symmetry. In $ \alpha $-I$_3$, the lattice effect is weak because the lower structural symmetry (even without lattice distortion at high temperatures) already ensures some charge disproportionation. 
Namely, site B is hole-rich and site C is hole-poor in Fig.~\ref{fig:structure}, which is regarded as a precursor to the horizontal-stripe charge order. Consequently, the electron-lattice coupling merely breaks the residual symmetry at low temperatures. Therefore, different tendencies of these salts to the transition to a metallic phase are reasonably expected 
and preliminarily reported by using the time-dependent Hartree-Fock approximation. \cite{yonemitsu_jpcs09,tanaka_jpcs09} 

In the present study, we employ the exact many-electron wave functions coupled with classical phonons in order to elucidate the roles of electron-lattice interactions and different structural symmetries for the photoinduced melting dynamics. Photoexcitation is introduced by an oscillating electric field, and the time evolution is obtained by solving the time-dependent Schr\"odinger equation. 
\cite{yonemitsu_prb07,onda_prl08} The absorbed energy that is required for 
the transition to a metallic phase 
is indeed larger for $ \theta $-RbZn than for $ \alpha $-I$_3$ indicating the 
higher stability 
of the charge order in $ \theta $-RbZn. Different structural symmetries lead to different charge dynamics through different degrees of degeneracy among charge ordering patterns. 

In reality, degrees of freedom outside the present model might affect the photoinduced dynamics. The differences between these salts are not limited to the conduction layers. Counter ions with different networks might influence such nonlinear dynamics. In the present work, however, we focus on the conduction layers within two-dimensional models. Numerical results will be useful for understanding and for 
designing ultrafast and controllable photoinduced dynamics in future. 

\section{Extended Peierls-Hubbard Model on Triangular Lattice}

For the quasi-two-dimensional organic conductors, we use the following two-dimensional 3/4-filled extended Hubbard model with electron-lattice couplings that modulate transfer integrals on the anisotropic triangular lattice shown in Fig.~\ref{fig:structure}, 
\begin{eqnarray}
H & = & 
\sum_{\langle ij \rangle \sigma} \left[
(t_{i,j} \pm \alpha_{i,j} u_{i,j}) 
e^{\mathrm{i}(e/\hbar c) \mbox{\boldmath $ \delta $}_{i,j} \cdot \mbox{\boldmath $ A $}(t)}
c^\dagger_{i\sigma} c_{j\sigma} +\mbox{H.c.}
\right]
\nonumber \\ & & 
+U\sum_i n_{i\uparrow} n_{i\downarrow}
+\sum_{\langle ij \rangle} V_{i,j} n_i n_j
+\sum_{\langle ij \rangle } \frac{K_{i,j}}{2} u_{i,j}^2
\nonumber \\ & & 
+\sum_{\langle ij \rangle } \frac{K_{i,j}}{2\omega_{i,j}^2} \dot{u}_{i,j}^2
\;, \label{eq:model}
\end{eqnarray}
where $ c^\dagger_{i\sigma} $ creates an electron with spin $ \sigma $ at site $ i $, $ n_{i\sigma} $=$ c^\dagger_{i\sigma} c_{i\sigma} $, and $ n_i $=$ \sum_\sigma n_{i\sigma} $. The parameter $ t_{i,j} $ denotes the transfer integral for the bond between the neighboring $ i $th and $ j $th sites. Schematic illustrations for the low-temperature structure of the conduction layer in $ \theta $-RbZn \cite{watanabe_jpsj04} and that in $ \alpha $-I$_3$ \cite{kakiuchi_jpsj07} are shown in Figs.~\ref{fig:structure}(a) and \ref{fig:structure}(b), respectively. In this paper, we use the transfer integrals estimated from the extended H\"uckel calculations based on the X-ray structural analyses \cite{watanabe_jpsj04,kakiuchi_jpsj07} at high temperatures $ t_{i,j} $=$ t_{i,j}^\mathrm{HT} $ as before, \cite{miyashita_prb07,miyashita_jpsj08} $ t_{c}^\mathrm{HT} $(=$ t_{c1}^\mathrm{HT} $=$ t_{c2}^\mathrm{HT} $)=0.035, $ t_{p1}^\mathrm{HT} $(=$ t_{p3}^\mathrm{HT} $=$ -t_{p2}^\mathrm{HT} $)=$ -t_{p4}^\mathrm{HT} $=0.095 for $ \theta $-RbZn and $ t_{a1}^\mathrm{HT} $=$-$0.0350, $ t_{a2}^\mathrm{HT} $=$-$0.0461, $ t_{a3}^\mathrm{HT} $=0.0181, $ t_{b1}^\mathrm{HT} $=0.1271, $ t_{b2}^\mathrm{HT} $=0.1447, $ t_{b3}^\mathrm{HT} $=0.0629, and $ t_{b4}^\mathrm{HT} $=0.0245 for $ \alpha $-I$_3$. Hereafter, we take eV as the unit of energy. The parameter $ U $ denotes the on-site Coulomb repulsion strength, and $ V_{i,j} $=$ V_c $ or $ V_p $ the intersite Coulomb repulsion strength between the neighboring $ i $th and $ j $th sites on the vertical or diagonal bonds, respectively. We apply the values $ U $=0.7, $ V_c $=0.35, and $ V_p $=0.3 to the interaction strengths for both salts, which reasonably reproduce the observed optical conductivity of $ \theta $-RbZn. \cite{miyashita_prb07} 

The quantity $ u_{i,j} $ denotes the $ i $th molecular displacement from the high-temperature structure, $ \dot{u}_{i,j} $ its time derivative, $ \alpha_{i,j} $, $ K_{i,j} $, and $ \omega_{i,j} $ the corresponding coupling constant, elastic coefficient, and bare phonon energy, respectively. The modulations of the transfer integrals from the high- to low-temperature phases are obtained by appropriate strengths of electron-lattice couplings, which reproduce the observed charge order with the help of Coulomb repulsion. \cite{tanaka_jpsj07,miyashita_prb07,tanaka_jpsj08,miyashita_jpsj08} For the values of appropriate strengths, see ref.~\citen{miyashita_prb07} for $ \theta $-RbZn and ref.~\citen{miyashita_jpsj08} for $ \alpha $-I$_3$. Because each modulation is a monotonic function of the corresponding coupling strength, the latter is uniquely determined by setting the former at the experimentally observed value, i.e., the value estimated from the extended H\"uckel calculation based on the X-ray structural analysis at low temperature. The signs of $ (t_{i,j} \pm \alpha_{i,j} u_{i,j}) $ in the first line of eq.~(\ref{eq:model}) are chosen so that $ y_{i,j} > 0 $ corresponds to the experimentally observed ones. For details, see ref.~\citen{miyashita_prb07} for $ \theta $-RbZn and ref.~\citen{miyashita_jpsj08} for $ \alpha $-I$_3$. For simplicity in our present calculations, however, we take only the most important electron-lattice coupling, $ s_\phi $ \cite{tanaka_jpsj07} equivalent to $ s_{p4} \equiv \alpha_{p4}^2/K_{p4} $ \cite{miyashita_prb07} that makes $ t_{p2} $ different from $ t_{p4} $ in $ \theta $-RbZn, 
$$ t_{p2} = t_{p4}^\mathrm{HT} + \alpha_{p4} u_{p4} $$ and 
$$ t_{p4} = t_{p4}^\mathrm{HT} - \alpha_{p4} u_{p4} \;, $$ 
and $ s_{b2} \equiv \alpha_{b2}^2/K_{b2} $ \cite{tanaka_jpsj08,miyashita_jpsj08} that makes $ t_{b2} $ different from $ t_{b2'} $ in $ \alpha $-I$_3$, 
$$ t_{b2}  = t_{b2}^\mathrm{HT} - \alpha_{b2} u_{b2} $$ and 
$$ t_{b2'} = t_{b2}^\mathrm{HT} + \alpha_{b2} u_{b2} \;. $$ 
Hereafter, the variables are transformed as 
\begin{equation}
\alpha_{i,j} u_{i,j} = y_{i,j}
\;, \ \ \ \ 
\frac{\alpha_{i,j}^2}{K_{i,j}}=s_{i,j}
\;,
\end{equation}
for the bond index $(i,j)$=$ p $4 in $ \theta $-RbZn and for $(i,j)$=$ b $2 in $ \alpha $-I$_3$. Now, we consider only one mode of molecular displacements. Then, we set the bare phonon energy at $ \omega_{i,j} $=$ \omega_\mathrm{ph} $. 

We use the 12-site clusters with periodic boundary conditions shown in Fig.~\ref{fig:structure}. It should be noted that the 12-site clusters are tiled in a staggered manner to cover the anisotropic triangular lattice. As a consequence, the left-right symmetry is lost even if $ p1 $=$ p3 $ in $ \theta $-RbZn. The rectangle denotes a unit cell. The labeling of bonds corresponds to the crystal structure at low temperatures. Hole-rich sites (A and B in $ \theta $-RbZn and in 
$ \alpha $-I$_3$) are denoted by gray ellipses, while hole-poor sites (C, D in $ \theta $-RbZn and C, A' in $ \alpha $-I$_3$) by white ellipses. Transfer integrals modulated by electron-lattice couplings are denoted by different thickness. 

In the Peierls phase of the transfer integral, $ e^{i(e/\hbar c) \mbox{\boldmath $ \delta $}_{i,j} \cdot \mbox{\boldmath $ A $}(t)} $, $ e $ denotes the absolute value of the electronic charge, $ c $ the light velocity, $ \mbox{\boldmath $ \delta $}_{i,j} $ the position vector $ \mbox{\boldmath $ \delta $}_{i,j} $=$ \mbox{\boldmath $ r $}_{j} - \mbox{\boldmath $ r $}_{i} $, and $ \mbox{\boldmath $ A $}(t) $ the time-dependent vector potential, 
\begin{equation}
\mbox{\boldmath $ A $}(t) = 
-c \int_0^t \mathrm{d} t' \mbox{\boldmath $ E $}(t')
\;, \label{eq:vector_potential}
\end{equation}
for the time-dependent electric field $ \mbox{\boldmath $ E $}(t) $. The field is here assumed to be given by 
\begin{equation}
\mbox{\boldmath $ E $}(t) = \mbox{\boldmath $ E $}_\mathrm{ext} 
\theta(t) \theta(T_\mathrm{irr}-t) \sin \omega_\mathrm{ext}t
\;, \label{eq:electric_field}
\end{equation}
with amplitude $ \mbox{\boldmath $ E $}_\mathrm{ext} $, frequency $ \omega_\mathrm{ext} $, and pulse width $ T_\mathrm{irr} $=$ 2\pi N_\mathrm{ext} / \omega_\mathrm{ext} $ with integer $ N_\mathrm{ext} $. Here, $ \theta(t) $ is the Heaviside step function, $ \theta(t) $=1 for $ t > $0 and $ \theta(t) $=0 for $ t < $0. By setting the intermolecular distance along the $ c $-axis of $ \theta $-RbZn and that along the $ a $-axis of $ \alpha $-I$_3$ as the unit of length and $ e $ at unity, the amplitude of the electric field $ \mid \mbox{\boldmath $ E $}_\mathrm{ext} \mid $ is given in units of eV. 

The time-dependent Schr\"odinger equation for the exact many-electron wave function $ \mid \! \psi (t) \rangle $ is numerically solved by 
\begin{equation}
\mid \! \psi (t+dt) \rangle \simeq 
\exp \left[-\frac{ \mathrm{i} }{ \hbar } dt H \left(t+\frac{ dt }{ 2 }\right) \right]
\mid \! \psi (t) \rangle 
\;,
\end{equation}
where the time evolution operator is expanded as 
\begin{equation}
\exp \left[-\frac{ \mathrm{i} }{ \hbar } dt H \left(t+\frac{ dt }{ 2 }\right) \right]
 = 
\sum_{n=0}^{\infty} \frac{1}{n!} 
\left[-\frac{ \mathrm{i} }{ \hbar } dt H \left(t+\frac{ dt }{ 2 }\right) \right]^n 
\;,
\end{equation}
with time slice $ dt $=0.054 for $ \theta $-RbZn and $ dt $=0.050 for $ \alpha $-I$_3$ up to the order by which the deviation of $ \langle \psi (t) \! \mid \! \psi (t) \rangle $ from unity is less than 10$^{-12}$. The initial state $ \mid \! \psi (0) \rangle $ is set at the ground state. The classical equation of phonon motion is solved by the leapfrog method, where the force is derived from the Hellmann-Feynman theorem. The initial displacement $ y(0) $ is at the minimum of the adiabatic potential for the ground state, so that the initial velocity $ \dot{y}(0) $ is zero. 

\section{Conductivity Spectra}

\begin{figure}
\includegraphics[height=12cm]{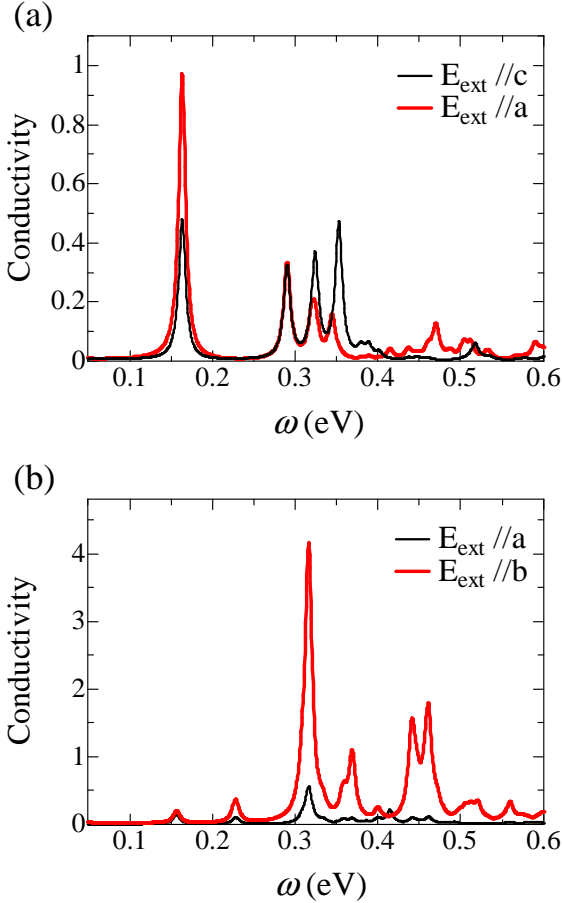}
\caption{(Color online) Optical conductivity spectra obtained by exact diagonalization of the 12-site clusters with $ U $=0.7, $ V_c $=0.35, and $ V_p $=0.3 for (a) $ \theta $-RbZn ($ s_{p4} $=0.1) and (b) $ \alpha $-I$_3$ ($ s_{b2} $=0.1) with different polarizations as indicated. The peak-broadening parameter $ \eta $ is set at 0.005. \label{fig:ED_absorption}}
\end{figure}

Before proceeding to the photoinduced dynamics, we calculate the regular parts of the optical conductivity spectra by the formula,
\begin{equation}
\sigma(\omega) = -\frac{1}{N\omega} \mathrm{Im} 
\left[
\langle \psi_0 \mid \mbox{\boldmath $ j $}
\frac{1}{\omega + \mathrm{i} \eta + E_0 - H}
\mbox{\boldmath $ j $} \mid \psi_0 \rangle
\right]
\;, \label{eq:conductivity}
\end{equation}
and show them in Fig.~\ref{fig:ED_absorption}.
Here, $ \mid \! \psi_0 \rangle $ denotes the ground state, $ E_0 $ its energy, $ \mbox{\boldmath $ j $} $ the current operator with rigid transfer integrals, $ \eta $ the peak-broadening parameter, and $ N $ the system size. Namely, the infrared conductivity originating from phonons is ignored. These spectra corresponding to $ \omega_\mathrm{ph} $=0 are useful for judging whether or not the frequency of the oscillating electric field (the energy of the pump light) is resonant with any charge-transfer excitation. In both cases of $ \theta $-RbZn and of $ \alpha $-I$_3$, the oscillator strengths are larger for polarization parallel to the stripes ($ a $- and $ b $-axes, respectively) than for polarization perpendicular to them ($ c $- and $ a $-axes, respectively). This is because the transfer integrals are larger for charge transfers along the $ p $- and $ b $-bonds, respectively, than those along the $ c $- and $ a $-bonds, respectively. 

The charge-transfer excitation with the largest oscillator strength in $ \theta $-RbZn [Fig.~\ref{fig:ED_absorption}(a)] has a lower energy and a smaller oscillator strength than that in $ \alpha $-I$_3$ [Fig.~\ref{fig:ED_absorption}(b)]. For comparison, we will use similar excitation energies $ \omega_\mathrm{ext} $ for the two compounds, which are almost resonant with excited states with substantial oscillator strengths. The targeted excited state with $ \omega $=0.2903 in $ \theta $-RbZn has charge densities, $ 2- \langle n_i \rangle $, 0.67 at sites A and B, and 0.33 at sites C and D, while that with $ \omega $=0.3162 in $ \alpha $-I$_3$ has 0.693 at site A, 0.648 at site B, 0.326 at site C, and 0.333 at site A'. Note that the ground state in $ \theta $-RbZn has charge densities, 0.93 at sites A and B, and 0.07 at sites C and D, while that in $ \alpha $-I$_3$ has 0.91 at site A, 0.78 at site B, 0.14 at site C, and 0.17 at site A'. Because of the smallness of the present clusters, these values themselves are insignificant, but they are useful when we analyze the respective photoinduced dynamics later. Concerning the overall shapes of the conductivity spectra, the time-dependent Hartree-Fock approximation for much larger system sizes gives smaller differences between $ \sigma(\omega) $ for $ \theta $-RbZn and $ \sigma(\omega) $ for $ \alpha $-I$_3$ than shown here. \cite{tanaka_unpub} The experimentally observed spectra show very broad features for both salts, \cite{dressel_cr04} so that a comparison is not straightforward. 

\section{Photoinduced Melting Dynamics}

For photoexcitations hereafter, we use the polarization parallel to the stripes and similar energies that are nearly resonant with a charge-transfer excitation with substantial oscillator strength. The action of photoexcitations is basically to destroy the charge order, so that excitation-density-dependent quantities are insensitive to the polarization and to 
the frequency of the oscillating electric field. \cite{yonemitsu_prb07}

\begin{figure}
\includegraphics[height=12cm]{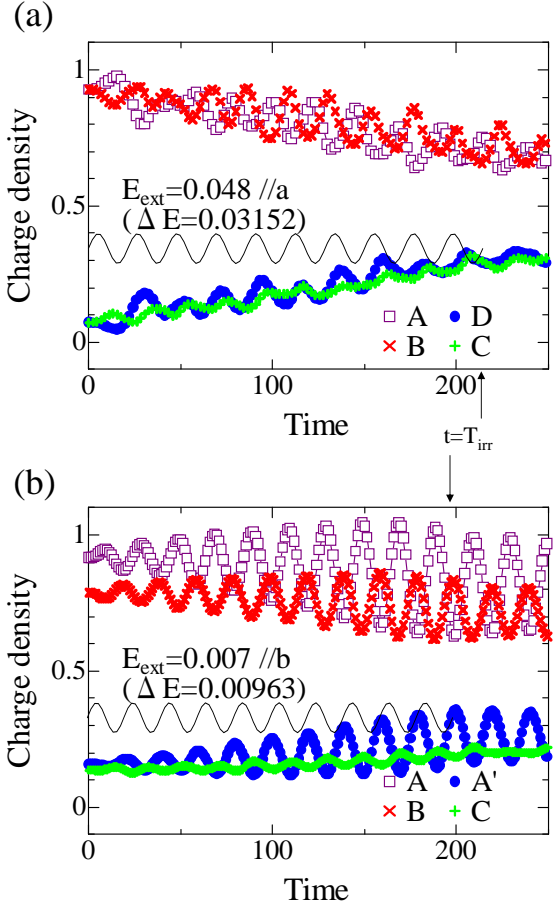}
\caption{(Color online) Time dependence of charge densities during ($ t < T_\mathrm{irr} $) and after ($ t > T_\mathrm{irr} $) photoexcitation along the stripes for (a) $ \theta $-RbZn ($ \omega_\mathrm{ext} $=0.2905, $ E_\mathrm{ext} $=0.048) and (b) $ \alpha $-I$_3$ ($ \omega_\mathrm{ext} $=0.3165, $ E_\mathrm{ext} $=0.007). The parameters shown in Fig.~\ref{fig:ED_absorption} and $ \omega_\mathrm{ph} $=0.01 are used. The thin lines show the time dependence of the electric field, eq.~(\ref{eq:electric_field}). 
Its period, $ 2\pi / \omega_\mathrm{ext} $, is 21.6 in (a) and 19.9 in (b).
\label{fig:ED_evolution}}
\end{figure}

The time evolution of the charge densities, $ 2- \langle n_i \rangle $, during ($ t < T_\mathrm{irr} $) and after ($ t > T_\mathrm{irr} $) photoexcitation is shown in Fig.~\ref{fig:ED_evolution}. 
Because we take eV as the unit of energy, $ t $=1520 corresponds to 1 ps. In both cases of $ \theta $-RbZn and of $ \alpha $-I$_3$, the amplitudes of the oscillating electric fields are chosen so that they are near critical values for melting the charge orders at $ t $=$ T_\mathrm{irr} $ with $ \omega_\mathrm{ext} $ shown in the caption and with 
$ N_\mathrm{ext} $=10. The total energy per site is increased by 0.03152 for $ \theta $-RbZn and by 
0.00963 for $ \alpha $-I$_3$. The resonantly-excited charge-transfer processes are clearly visible, which have periods that are nearly equal to those of the oscillating fields $ 2\pi / \omega_\mathrm{ext} $ [21.6 in Fig.~\ref{fig:ED_evolution}(a) and 19.9 in Fig.~\ref{fig:ED_evolution}(b)]. In the figure, the time dependence of the electric field, eq.~(\ref{eq:electric_field}), is shown by the thin lines for comparison. The amplitudes of such oscillating charge densities with these frequencies are slightly larger for $ \alpha $-I$_3$. There is a nonnegligible difference in the overall time dependences. The oscillation for $ \theta $-RbZn is more complex [Fig.~\ref{fig:ED_evolution}(a)], while that for $ \alpha $-I$_3$ consists of a fewer components [Fig.~\ref{fig:ED_evolution}(b)]. We always find such difference irrespective of polarization or energy of the photoexcitation (not shown). Such difference is also seen by the time-dependent Hartree-Fock approximation for much larger system sizes, \cite{tanaka_unpub} so that systematic differences seem to exist in the electronic states. The origins of these behaviors are discussed below. 

To analyze the character of the electronic wave function at time $ t $, $ \mid \! \psi(t) \rangle $, we can expand it by using some orthonormal functions. 
Because the applied field is uniform, the molecular displacements $ y_{i,j}(t) $ are homogeneous, $ y_{i,j}(t) = y(t) $, which is, of course, a classical variable and is obtained as the solution to the classical equation of phonon motion. For each $ y(t) $, we employ the eigenfunctions of the Hamiltonian, $ \mid \! \psi_j[ y(t) ] \rangle $, for such orthonormal functions. 
At $ t $=0, $ \mid \! \psi(0) \rangle $ is set at the ground state $ \mid \! \psi_0[ y(0) ] \rangle $. At $ t > $0, $ \mid \! \psi(t) \rangle $ deviates from 
the lowest-energy state 
$ \mid \! \psi_0[ y(t) ] \rangle $ 
of the Hamiltonian with $ y(t) $. 
Then, let us define the ground-state population by the square of the absolute value of the overlap between the wave function, $ \mid \! \psi(t) \rangle $, and the lowest-energy 
state $ \mid \! \psi_0[ y(t) ] \rangle $ for the molecular displacement $ y(t) $. 
Technically, we obtain this quantity by projecting the wave function $ \mid \! \psi(t) \rangle $ to $ \mid \! \psi_0[ y(t) ] \rangle $ at each step of time evolution. 
If the time evolution is adiabatic, this quantity is independent of time. 

\begin{figure}
\includegraphics[height=12cm]{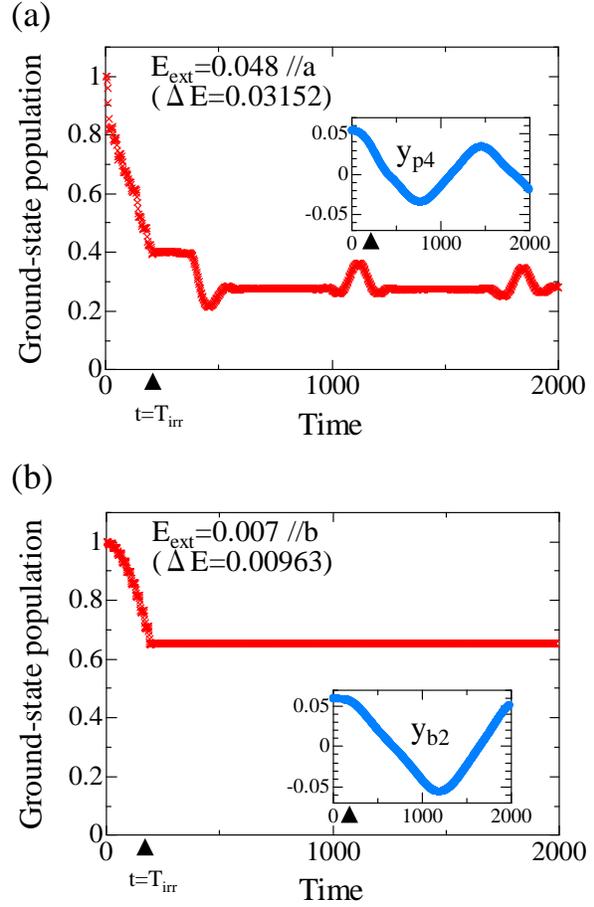}
\caption{(Color online) Time dependence of ground-state populations $ \mid \langle \psi_0[ y(t) ] \mid \psi(t) \rangle \mid^2 $ and modulations of transfer integrals $ y $ (inset) during and after photoexcitation along the stripes for (a) $ \theta $-RbZn with $ y $=$ y_{p4} $ and (b) $ \alpha $-I$_3$ with $ y $=$ y_{b2} $. The parameters shown in Fig.~\ref{fig:ED_evolution} are used. \label{fig:ED_population}}
\end{figure}

The time dependence of the ground-state population 
is shown in Fig.~\ref{fig:ED_population}. 
The pulse width $ T_\mathrm{irr} $ is denoted by the triangles. In $ \theta $-RbZn, the ground-state population is about 0.4 at $ t $=$ T_\mathrm{irr} $ and shows a complex behavior for $ t > T_\mathrm{irr} $ [Fig.~\ref{fig:ED_population}(a)]. 
It varies significantly when the displacement $ y_{p4}(t) $ is small. 
As discussed in the following paragraph, the undistorted ($ y_{i,j} $=0) lattice structure possesses the high symmetry in $ \theta $-RbZn, leading to quasi-degeneracy of excited states with different charge distributions. The weights of these excited states in $ \mid \! \psi(t) \rangle $ are sensitively altered by $ y_{p4}(t) $. This causes the variation of the ground-state population near $ y_{p4}(t) $=0. In $ \alpha $-I$_3$, on the other hand, the ground-state population is about 0.65 at $ t $=$ T_\mathrm{irr} $, which is larger than in $ \theta $-RbZn, and becomes independent of time for $ t > T_\mathrm{irr} $ [Fig.~\ref{fig:ED_population}(b)]. In contrast to $ \theta $-RbZn, the undistorted ($ y_{i,j} $=0) lattice structure possesses quite low symmetry in $ \alpha $-I$_3$, which is free from such (quasi-)degeneracy. This causes a constant ground-state population for $ t > T_\mathrm{irr} $. The ground-state populations at $ t $=$ T_\mathrm{irr} $ are rather insensitive to the polarization and to 
the energy of the photoexcitation. For any combination of the polarization and the energy, 
the ground-state population at $ t $=$ T_\mathrm{irr} $ 
for $ \theta $-RbZn is smaller than that for $ \alpha $-I$_3$. This indicates that a smaller number of excited states are involved in the dynamics for $ \alpha $-I$_3$. 
The oscillations of the charge densities shown in Fig.~\ref{fig:ED_evolution}(b) are nearly sinusoidal because only a few excited states are involved in the photoinduced dynamics for $ \alpha $-I$_3$. 
Both of the displacements $ y_{p4} $ and $ y_{b2} $ show an oscillator's behavior in the insets of Fig.~\ref{fig:ED_population}. The periods of the oscillations are about 1500 to 2000, which are much larger than those of bare phonons ($ 2\pi/\omega_\mathrm{ph} $=628) owing to the electron-lattice interaction (phonon softening). The oscillators seem to be weakly damped, but they do not vanish within accessible computation times. 

\begin{figure}
\includegraphics[height=12cm]{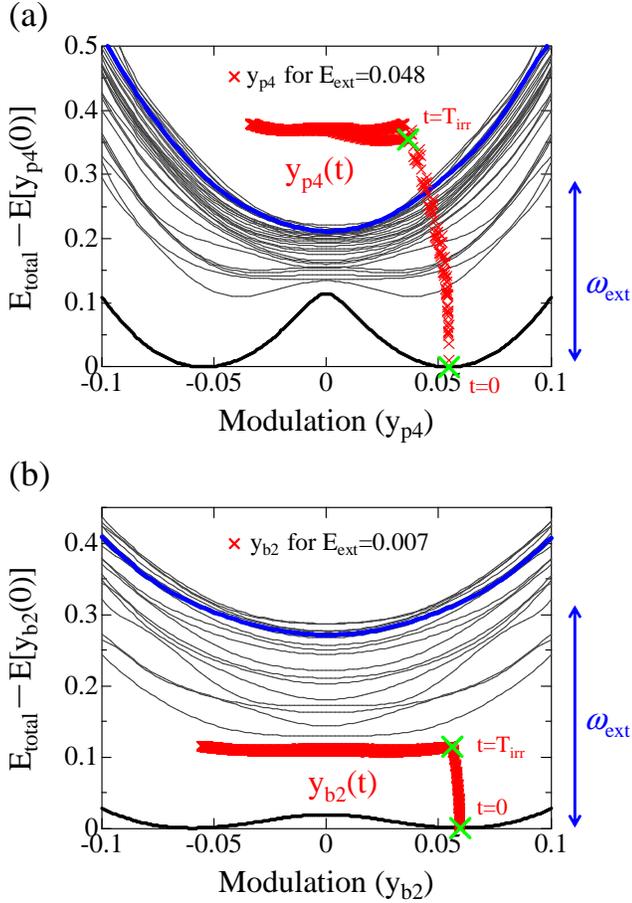}
\caption{(Color online) Adiabatic potentials of ground and excited states for (a) $ \theta $-RbZn  and (b) $ \alpha $-I$_3$. Those of the resonantly-excited charge-transfer states are presented in blue (by thick lines). The trajectories of the electronic energies with $ y_{p4}(t) $ and $ y_{b2}(t) $ are plotted in red (by crosses) for the parameters used in Fig.~\ref{fig:ED_evolution}. \label{fig:ED_trajectory}}
\end{figure}

For the sake of discussions later, let us set aside the time evolution for the moment and calculate the total energies for the eigenstates  $ \mid \! \psi_j[ y(t) ] \rangle $ on the assumption of $ \dot{y}=0 $. 
The adiabatic potentials for the ground and excited states 
obtained in this manner 
are shown in Fig.~\ref{fig:ED_trajectory}, as a function of the molecular displacement $ y_{p4} $ or $ y_{b2} $. 
For both $ \theta $-RbZn and $ \alpha $-I$_3$, there are resonantly-excited charge-transfer states in Figs.~\ref{fig:ED_evolution} and \ref{fig:ED_population}, which have substantial oscillator strengths and excitation energies of about $ \omega_\mathrm{ext} \simeq $0.3 (Fig.~\ref{fig:ED_absorption}) at the potential minima for the ground states (i.e., at $ y \simeq $0.05). They are indicated in blue (by thick lines). 
For excited states with 
much higher energies than those of the resonantly-excited charge-transfer states, 
their adiabatic potentials are omitted for the sake of clarity. First of all, the lattice stabilization energy is larger for $ \theta $-RbZn [0.114 in total or 0.0095 per site in Fig.~\ref{fig:ED_trajectory}(a)] than for $ \alpha $-I$_3$ [0.019 in total or 0.0016 per site in Fig.~\ref{fig:ED_trajectory}(b)]. In $ \theta $-RbZn, the energies are distributed more densely at $ y_{p4} $=0 than in $ \alpha $-I$_3$. It is because different patterns of charge orders are nearly degenerate owing to the high symmetry of its undistorted lattice structure and the resultant charge frustration. \cite{miyashita_prb07} 

Now, we return to the analysis of the photoinduced dynamics. Here we show 
the trajectory of the electronic energy, which is obtained by subtracting the lattice kinetic energy, 
$ \sum_{\langle ij \rangle } (K_{i,j}/2\omega_{i,j}^2) \dot{u}_{i,j}^2 $, 
from the total energy. It should be noted again that, for both $ \theta $-RbZn and $ \alpha $-I$_3$, the amplitudes of the oscillating electric fields, $ E_\mathrm{ext} $, are chosen so that they are near critical values for melting the charge orders. The total energy is of course conserved after photoexcitation ($ t > T_\mathrm{irr} $). The electronic energy and the lattice kinetic energy are not monotonic functions of time because some energy is transferred back and forth. 
The electronic energy after photoexcitation ($ t > T_\mathrm{irr} $) is not a symmetric function of $ y $. It is clearly seen for $ \theta $-RbZn in Fig.~\ref{fig:ED_trajectory}(a), where the ground-state population shows a complex behavior [Fig.~\ref{fig:ED_population}(a)]. 
Quantitatively, the lattice kinetic energy is small in the present model, which restricts the number of phonon modes to one. 
In general, with an increasing number of phonon modes that are coupled to the electronic system, more energy would be transferred to the phonon system. However, the electronic energy dominates the total energy because the nearest-neighbor repulsive interactions are much larger than the electron-lattice interactions. 

The electronic energy reaches an energy much higher than that of the resonantly-excited charge-transfer state for $ \theta $-RbZn [Fig.~\ref{fig:ED_trajectory}(a)]. This indicates that quite a large number of excited states are involved in the photoinduced dynamics. In $ \alpha $-I$_3$, on the other hand, the energies are distributed sparsely around $ y_{b2} $=0 owing to the low symmetry even in the undistorted structure. Recall that charge disproportionation already exists between sites B and C at $ y_{b2} $=0, and a finite $ y_{b2} $ breaks the residual symmetry between sites A and A'. The trajectory of the electronic energy reaches an energy that is much lower than that of the resonantly-excited charge-transfer state for $ \alpha $-I$_3$ [Fig.~\ref{fig:ED_trajectory}(b)]. This indicates that a much smaller number of excited states are involved in this photoinduced dynamics. 

\section{Excitation-Density-Dependent Correlations}

Hereafter, charge densities and correlation functions are calculated with different amplitudes, $ E_\mathrm{ext} $. For each $ E_\mathrm{ext} $, they are averaged after $ t $=$ T_\mathrm{irr} $ over the period of $ T_\mathrm{irr} $=$ 2\pi N_\mathrm{ext} / \omega_\mathrm{ext} < t < 2\pi N_\mathrm{obs} / \omega_\mathrm{ext} $ with $ N_\mathrm{ext} $=10 and $ N_\mathrm{obs} $=100. They are shown as a function of the increment in the total energy per site $ \Delta E $. 

\begin{figure}
\includegraphics[height=12cm]{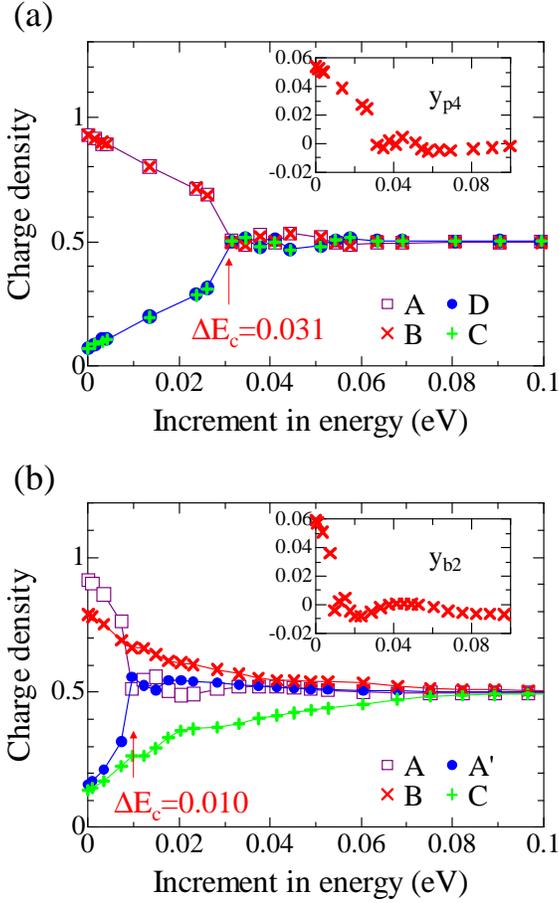}
\caption{(Color online) Time averages of charge densities and modulations of transfer integrals (inset) after photoexcitation along the stripes for (a) $ \theta $-RbZn ($ \omega_\mathrm{ext} $=0.2905) and (b) $ \alpha $-I$_3$ ($ \omega_\mathrm{ext} $=0.3165), as a function of increment in total energy  $ \Delta E $. The parameters shown in Fig.~\ref{fig:ED_absorption} and $ \omega_\mathrm{ph} $=0.01 are used. \label{fig:ED_density}}
\end{figure}

The charge densities, $ 2- \langle n_i \rangle = \langle n^h_i \rangle $ with $ n^h_i $=$ \sum_\sigma n^h_{i\sigma} $, $ n^h_{i\sigma} $=$ c^{h\dagger}_{i\sigma} c^h_{i\sigma} $, and $ c^h_{i\sigma} $=$ c^\dagger_{i\sigma} $, at four sites 
in the unit cell of Fig.~\ref{fig:structure} 
after photoexcitation are plotted in Fig.~\ref{fig:ED_density}. 
In each case, there is a critical value of the increment in the total energy per site $ \Delta E_c $, above which the interaction-induced charge disproportionation disappears. The critical value [$ \Delta E_c $=0.031 in Fig.~\ref{fig:ED_density}(a)] for $ \theta $-RbZn is larger than that [$ \Delta E_c $=0.010 in Fig.~\ref{fig:ED_density}(b)] for $ \alpha $-I$_3$. This is understandable in light of the fact that the lattice stabilization energy is larger for $ \theta $-RbZn. We have calculated the critical values with different polarizations and energies of photoexcitations. For the polarization perpendicular to the stripes and for 
the same excitation energies as in this section, the critical value for $ \theta $-RbZn is $ \Delta E_c $=0.020 and that for $ \alpha $-I$_3$ is $ \Delta E_c $=0.008. For the polarization parallel to the stripes and for 
$ \omega_\mathrm{ext} $=0.25, which is far from any charge-transfer excitation energy (Fig.~\ref{fig:ED_absorption}), the critical value for $ \theta $-RbZn is $ \Delta E_c $=0.028 and that for $ \alpha $-I$_3$ is $ \Delta E_c $=0.014. We always find the relation, 
$ \Delta E_c $ for $ \theta $-RbZn is larger than $ \Delta E_c $ for $ \alpha $-I$_3$. 

Above $ \Delta E_c $, the time-averaged charge densities are uniform for $ \theta $-RbZn, but this is not the case for $ \alpha $-I$_3$ with low structural symmetry. For $ \alpha $-I$_3$, the time-averaged charge density at site A and that at site A' are equal above $ \Delta E_c $. The difference between those at sites B and C decreases with increasing photoabsorption, and it finally vanishes after photoabsorption much larger than $ \Delta E_c $. The insets of Fig.~\ref{fig:ED_density} show that the molecular displacements that modulate the transfer integrals also vanish around $ \Delta E_c $. In both cases, the electronic states above $ \Delta E_c $ are regarded as metallic because of almost vanishing 
charge correlations shown later. 

\begin{figure}
\includegraphics[height=12cm]{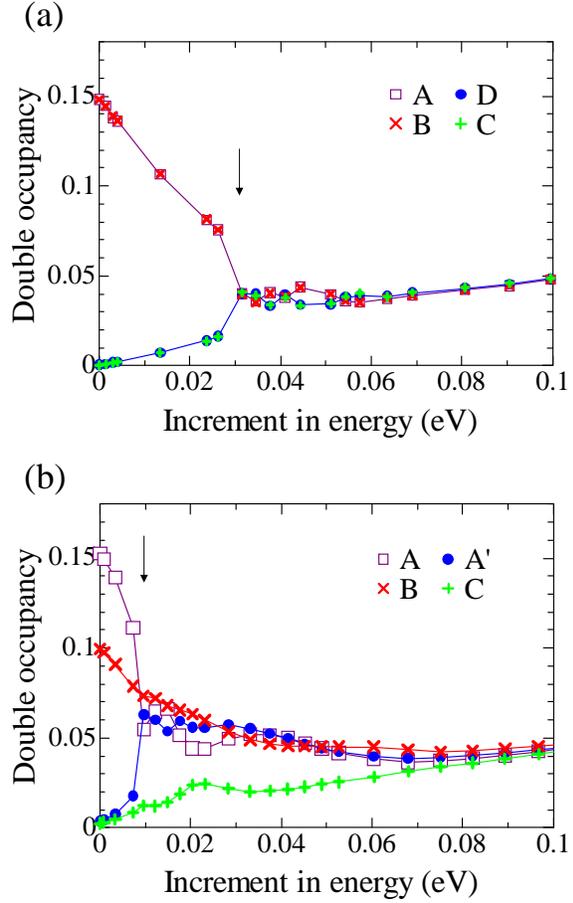}
\caption{(Color online) Time averages of double occupancies after photoexcitation along the stripes for (a) $ \theta $-RbZn ($ \omega_\mathrm{ext} $=0.2905) and (b) $ \alpha $-I$_3$ ($ \omega_\mathrm{ext} $=0.3165), as a function of increment in total energy $ \Delta E $. $ \Delta E_c $ is indicated by the arrows. The parameters shown in Fig.~\ref{fig:ED_absorption} and $ \omega_\mathrm{ph} $=0.01 are used. \label{fig:ED_double}}
\end{figure}

The double occupancies, $ \langle n^h_{i\uparrow} n^h_{i\downarrow} \rangle $, after photoexcitation are plotted in Fig.~\ref{fig:ED_double}. 
Their behavior is quite similar to that of the charge densities. Above the same critical values of $ \Delta E_c $, the interaction-induced differences in the double occupancies disappear. In $ \theta $-RbZn, the double occupancy is large at hole-rich sites and very small at hole-poor sites below $ \Delta E_c $ [Fig.~\ref{fig:ED_double}(a)]. The difference becomes small with increasing photoabsorption and disappears at $ \Delta E_c $. Above $ \Delta E_c $, the time-averaged double occupancies are uniform and increase with photoabsorption for $ \theta $-RbZn. For $ \alpha $-I$_3$, that at site A and that at site A' are equal above $ \Delta E_c $ [Fig.~\ref{fig:ED_double}(b)]. The difference between those at sites B and C decreases with increasing photoabsorption, and it finally vanishes after photoabsorption much larger than $ \Delta E_c $. In both cases, the double occupancies approach 0.0625=(1/4)$^2$ for very large photoabsorption, which is the value the noninteracting uniform state possesses. 

\begin{figure}
\includegraphics[height=11cm]{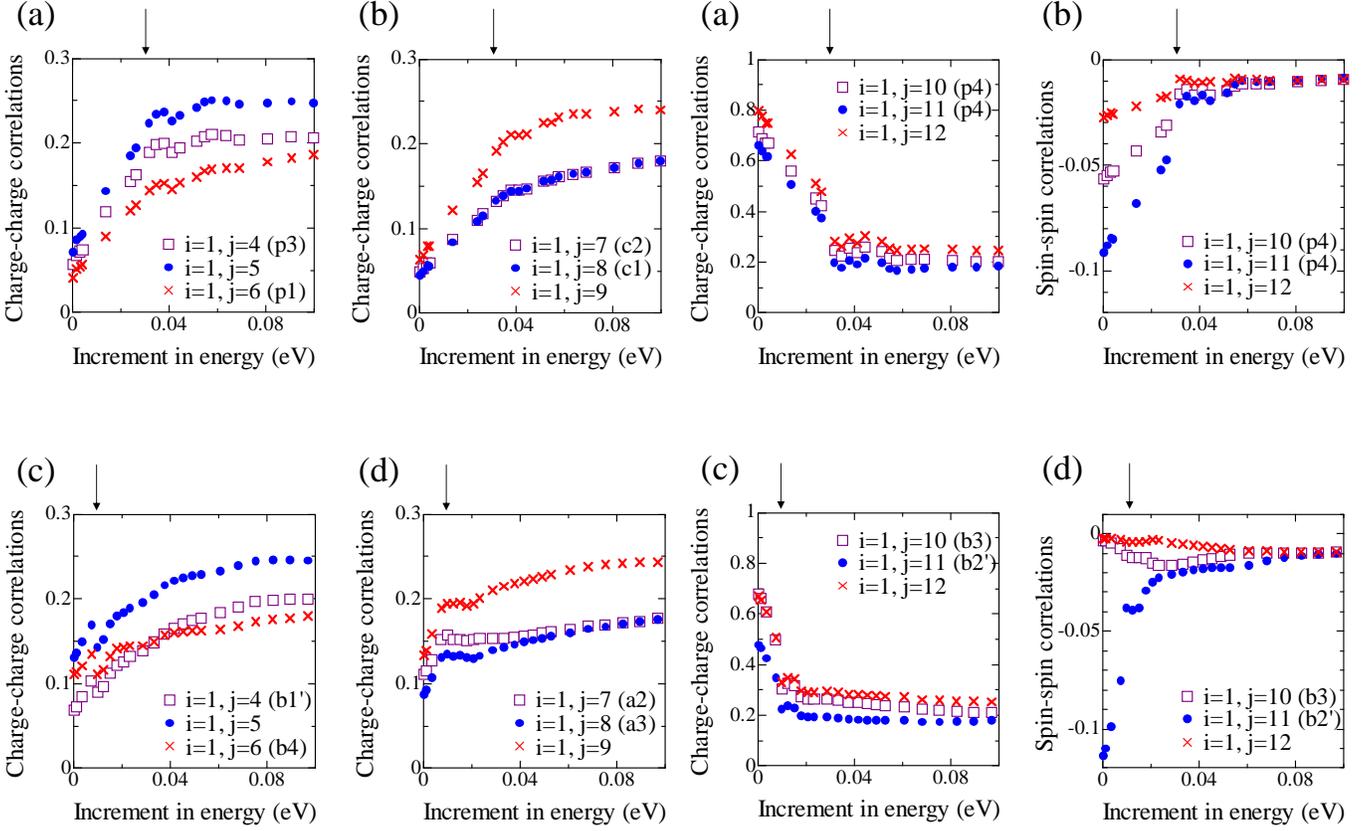}
\caption{(Color online) Time averages of charge-charge correlation functions between hole-rich and hole-poor sites after photoexcitation along the stripes, (a-b) for $ \theta $-RbZn ($ \omega_\mathrm{ext} $=0.2905) and (c-d) for $ \alpha $-I$_3$ ($ \omega_\mathrm{ext} $=0.3165), as a function of increment in total energy $ \Delta E $. $ \Delta E_c $ is indicated by the arrows. The parameters shown in Fig.~\ref{fig:ED_absorption} and $ \omega_\mathrm{ph} $=0.01 are used. \label{fig:ED_charge_corr}}
\end{figure}

The charge-charge correlation functions, $ \langle n^h_i n^h_j \rangle $, with $ i $=1 on (hole-rich) site A and $ j $=4, 5, or 6 on (hole-poor) site C [Figs.~\ref{fig:ED_charge_corr}(a) and \ref{fig:ED_charge_corr}(c)] or $ j $=7, 8, or 9 on (hole-poor) site D [Fig.~\ref{fig:ED_charge_corr}(b)] or A' [Fig.~\ref{fig:ED_charge_corr}(d)] after photoexcitation are plotted here for $ \theta $-RbZn [Figs.~\ref{fig:ED_charge_corr}(a) and \ref{fig:ED_charge_corr}(b)] and for 
$ \alpha $-I$_3$ [Figs.~\ref{fig:ED_charge_corr}(c) and \ref{fig:ED_charge_corr}(d)]. 
Those between the nearest-neighbor sites are indicated in the graph legends by the labeling of bonds. These charge-charge correlation functions between hole-rich and hole-poor sites are small before photoexcitation (they are zero in the strong-coupling limit). They increase with photoabsorption below $ \Delta E_c $, and change only a little above $ \Delta E_c $. Between the distant pair of sites [for $ j $=5 in Figs.~\ref{fig:ED_charge_corr}(a) and \ref{fig:ED_charge_corr}(c), and for $ j $=9 in Figs.~\ref{fig:ED_charge_corr}(b) and \ref{fig:ED_charge_corr}(d)], it is near 0.25=(1/2)$^2$ above $ \Delta E_c $, which is the value the noninteracting uniform state possesses. In other words, the charge-charge correlations almost disappear 
above $ \Delta E_c $, indicating that these states are metallic with photocarriers. 

\begin{figure}
\includegraphics[height=11cm]{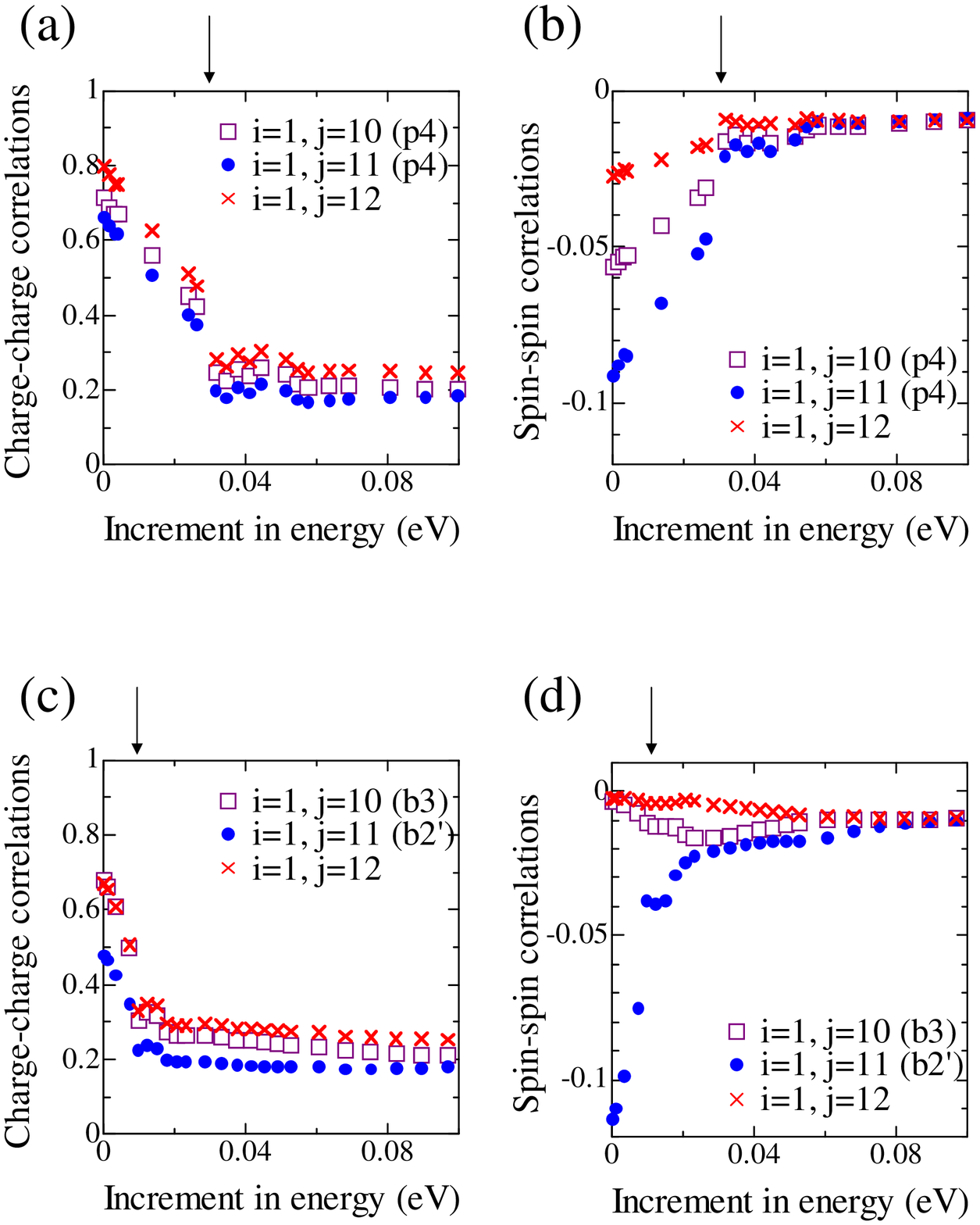}
\caption{(Color online) Time averages of (a,c) charge-charge correlation functions and (b,d) spin-spin correlation functions between hole-rich sites after photoexcitation along the stripes, (a-b) for $ \theta $-RbZn ($ \omega_\mathrm{ext} $=0.2905) and (c-d) for $ \alpha $-I$_3$ ($ \omega_\mathrm{ext} $=0.3165), as a function of increment in total energy  $ \Delta E $. $ \Delta E_c $ is indicated by the arrows. The parameters shown in Fig.~\ref{fig:ED_absorption} and $ \omega_\mathrm{ph} $=0.01 are used. \label{fig:ED_HRbond_corr}}
\end{figure}

The charge-charge correlation functions with $ i $=1 and $ j $=10, 11, or 12 between hole-rich sites A and B [Figs.~\ref{fig:ED_HRbond_corr}(a) and \ref{fig:ED_HRbond_corr}(c)] and the spin-spin correlation functions, $ \langle S^z_i S^z_j \rangle $ with $ S^z_i = (n_{i\uparrow}-n_{i\downarrow})/2 $, for the same combinations of $ i $=1 and $ j $ [Figs.~\ref{fig:ED_HRbond_corr}(b) and \ref{fig:ED_HRbond_corr}(d)] after photoexcitation are plotted here for $ \theta $-RbZn [Figs.~\ref{fig:ED_HRbond_corr}(a) and \ref{fig:ED_HRbond_corr}(b)] and for 
$ \alpha $-I$_3$ [Figs.~\ref{fig:ED_HRbond_corr}(c) and \ref{fig:ED_HRbond_corr}(d)]. 
The correlation functions between the nearest-neighbor sites are indicated in the graph legends by the labeling of bonds. The charge-charge correlation functions between hole-rich sites are large before photoexcitation (they are unity in the strong-coupling limit). They decrease with increasing photoabsorption below $ \Delta E_c $, and change only a little above $ \Delta E_c $. Between the distant pair of sites ($ i $=1 and $ j $=12), it is near 0.25=(1/2)$^2$ above $ \Delta E_c $ [Figs.~\ref{fig:ED_HRbond_corr}(a) and \ref{fig:ED_HRbond_corr}(c)], which is the value the noninteracting uniform state possesses. 

For $ \theta $-RbZn, the spin-spin correlation functions are negative and large on the $ p4 $ bonds (the difference between $ j $=10 and $ j $=11 is due to the loss of the left-right symmetry in the present tiling of the 12-site clusters) before photoexcitation [Fig.~\ref{fig:ED_HRbond_corr}(b)]. Their magnitudes decrease with increasing photoabsorption below $ \Delta E_c $, and they are very small and almost independent of the pair of sites above $ \Delta E_c $. Between the distant pair of sites ($ i $=1 and $ j $=12), it is always very small. For $ \alpha $-I$_3$, the spin-spin correlation functions are significant only on the $ b2' $ bonds before photoexcitation [Fig.~\ref{fig:ED_HRbond_corr}(d)]. Its magnitude decreases with increasing photoabsorption below $ \Delta E_c $, but it remains finite just above $ \Delta E_c $ due to the large $ t_{b2} $ even without lattice distortion (though it becomes very small for $ \Delta E $ much larger than $ \Delta E_c $). The other spin-spin correlation functions are always very small not only between the distant pair of sites but also on the $ b3 $ bonds. Just above $ \Delta E_c $, 
the spin-spin correlations almost disappear for $ \theta $-RbZn [Fig.~\ref{fig:ED_HRbond_corr}(b)], but they remain nearly spin-singlets on the $ b2' $ bonds for $ \alpha $-I$_3$ [Fig.~\ref{fig:ED_HRbond_corr}(d)]. This can be interpreted as follows. The energy transfer from the oscillating electric field to the spin degrees of freedom in the electronic state does (does not) reach 100\% in $ \theta $-RbZn ($ \alpha $-I$_3$). 

\section{Summary}

Differences between the photoinduced melting dynamics of the charge order in $ \theta $-(BEDT-TTF)$_2$RbZn(SCN)$_4$ ($ \theta $-RbZn) and that in $ \alpha $-(BEDT-TTF)$_2$I$_3$ ($ \alpha $-I$_3$) are studied in two-dimensional 3/4-filled extended Peierls-Hubbard models on anisotropic triangular lattices. The charge ordering patterns are very similar and known as horizontal stripes. They are stabilized by nearest-neighbor repulsive interactions and by 
electron-lattice interactions, but their relative importance in stabilizing the charge orders is different between these salts. \cite{tanaka_jpsj07,miyashita_prb07,tanaka_jpsj08,miyashita_jpsj08,tanaka_jpsj09} Time evolutions of exact many-electron wave functions coupled with classical phonons are obtained by solving the time-dependent Schr\"odinger equation during and after the application of an oscillating electric field. Calculations are performed for photoexcitations with different polarizations and energies. The qualitative difference between these dynamics is basically consistent with the experimentally observed one: \cite{iwai_prl07} the photoabsorption required for the transition to a metallic phase 
is larger for $ \theta $-RbZn than for $ \alpha $-I$_3$ indicating the higher stability 
of the charge order in $ \theta $-RbZn. 

Although the charge ordering patterns are very similar, these salts have different crystal structures. The high symmetry in the high-temperature structure of $ \theta $-RbZn causes charge frustration leading to quasi-degeneracy among electronic states possessing different charge ordering patterns if lattice distortions are absent. When the lattice distortions are small, the energy distribution is so dense that a large number of excited states are involved in the photoinduced dynamics. This causes a complex time dependence of the charge densities. The low symmetry in the high-temperature structure of $ \alpha $-I$_3$ causes a precursor to the charge order without lattice distortion. The energy distribution is rather sparse, and a fewer number of excited states are involved in the photoinduced dynamics. This causes a sinusoidal time dependence of the charge densities. 

Charge-charge and spin-spin correlation functions as well as charge densities and molecular displacements are averaged over time after photoexcitation and plotted as a function of the increment in the total energy per site $ \Delta E $. There is a critical value $ \Delta E_c $ in the increment, above which the interaction-induced charge disproportionation and lattice distortion disappear. The critical value $ \Delta E_c $ is larger for $ \theta $-RbZn. This relation always holds 
irrespective of polarization or energy of the photoexcitation. The correlation functions monotonically change with photoabsorption toward $ \Delta E_c $ from below. Their changes are small above $ \Delta E_c $, where the charge-charge correlations almost disappear 
and the system is regarded as metallic. Only on the bonds between hole-rich sites, the spin-spin correlations can be large before photoexcitation. Above $ \Delta E_c $, they become very weak and are still non-negligible only on the $ b2' $ bonds in $ \alpha $-I$_3$, where the transfer integrals are large even without lattice distortion owing to the low structural symmetry. 

From the theoretical viewpoint, the difference in the photoinduced dynamics of these salts is a quantitative one. The higher stability 
of the charge order in $ \theta $-RbZn is due to large stabilization energy. In reality, lattice anharmonicity might be important. For instance, the difference in the insulating layers of counter ions [RbZn(SCN)$_4$ in $ \theta $-RbZn and I$_3$ in $ \alpha $-I$_3$] might also contribute to the stability 
of the charge order in $ \theta $-RbZn compared with that in $ \alpha $-I$_3$. What is peculiar to $ \alpha $-I$_3$ is such low structural symmetry that charge disproportionation exists between sites B and C even in the metallic phase and the transition to the metallic phase 
is achieved by dissolving only the charge disproportionation between sites A and A'. 

\section*{Acknowledgment}

The authors are grateful to T. Yasuike for enlightening discussions. 
This work was supported by Grants-in-Aid for Scientific Research (C) (Grant No. 19540381), for Scientific Research (B) (Grant No. 20340101), and ``Grand Challenges in Next-Generation Integrated Nanoscience" from the Ministry of Education, Culture, Sports, Science and Technology of Japan, and NINS' Creating Innovative Research Fields Project (NIFS08KEIN0091).

\bibliography{pipt2dco}

\end{document}